\documentclass[12pt,preprint]{aastex}
\usepackage{graphicx}
\usepackage{float}

\def\xsrc{PSR~J1119$-$6127~}
\def\xsrcnos{PSR~J1119$-$6127}

\begin{document}
\vspace{0.8 in}

\title{Magnetar-like X-ray Bursts from a Rotation Powered Pulsar, \xsrc}

\author{Ersin G\"o\u{g}\"u\c{s}\altaffilmark{1},
Lin Lin \altaffilmark{2},
Yuki Kaneko\altaffilmark{1},
Chryssa Kouveliotou\altaffilmark{3},
Anna L. Watts\altaffilmark{4},
Manoneeta Chakraborty\altaffilmark{1},
M. Ali Alpar\altaffilmark{1},
Daniela Huppenkothen\altaffilmark{5},
Oliver J. Roberts\altaffilmark{6},
George Younes\altaffilmark{3},
Alexander J. van der Horst\altaffilmark{3}
}

\altaffiltext{1}{Sabanc\i~University, Orhanl\i$-$Tuzla, \.Istanbul 34956, Turkey}
\altaffiltext{2}{Department of Astronomy, Beijing Normal University, Beijing 100875, China}
\altaffiltext{3}{Department of Physics, The George Washington University, Washington, DC 20052, USA}
\altaffiltext{4}{Anton Pannekoek Institute for Astronomy, University of Amsterdam, Postbus 94249, NL-1090 GE Amsterdam, the Netherlands}
\altaffiltext{5}{Center for Data Science, New York University, 726 Broadway, 7th Floor, New York, NY 10003, USA; Center for Cosmology and Particle Physics, Department of Physics, New York University, 4 Washington Place, New York, NY 10003, USA}
\altaffiltext{6}{School of Physics, University College Dublin, Stillorgan Road, Belfield, Dublin 4, Ireland}

\begin{abstract}

Two energetic hard X-ray bursts have recently triggered the {\it Fermi} and {\it Swift} space observatories from the rotation powered pulsar, \xsrcnos. We have performed in depth spectral and temporal analyses of these two events. Our extensive searches in both observatory data for lower luminosity bursts uncovered 10 additional events from the source. We report here on the timing and energetics of the 12 bursts from \xsrc during its burst active phase of 2016 July 26 and 28. We also found a spectral softer X-ray flux enhancement in a post burst episode, which shows evidence of cooling. We discuss here the implications of these results on the nature of this unusual high-field radio pulsar, which firmly place it within the typical magnetar population.

\end{abstract}


\section{Introduction}

Episodic X-ray burst emission from magnetars has been heretofore attributed to diverse mechanisms associated with their extreme magnetic fields ($\sim$10$^{14}$$-$10$^{15}$ G). However, the detection of magnetar-like bursts from the young, rotation powered pulsar (RPP) PSR\,J$1846-0258$ (Gavriil et al 2008), and from a magnetar with a surprisingly low magnetic field, SGR\,$0418+5729$ ($6.1\times10^{12}$ G; similar to the typical surface dipole fields of ordinary RPPs; \citet{Rea2010}, see \citet{vdHorst2010} for bursts), suggested that the two populations maybe actually linked via a continuum of magnetic activity. 

Typical magnetar bursts are brief ($\sim0.1$\,s long) but very luminous, reaching peak luminosities of about $10^{41}$ erg/s \citep{Gogus2001,Gavriil2004,vdHorst2012,Younes2014}. These constitute the bulk of burst activity, with a few intermediate bursts of about an order of magnitude more energetic, longer durations, and long lasting tail emission, which is much weaker than the burst but significantly above the persistent emission level (Lenters et al. 2003, G{\"o}{\v g}{\"u}{\c s} et al. 2011, Chakraborty et al. 2016). 

Several mechanisms have been proposed as the source of magnetar bursts; they all assume that these are powered by their fields (for a review see Turolla, Zane \& Watts 2015). The crust quake model posits that the dissipation of internal magnetic energy strains the solid crust of the neutron star, which then fractures when the magnetic pressure on it becomes larger than the limiting stress it could resist. This is followed by particle acceleration and emission of radiation in the form of a short burst (Thompson \& Duncan 1995). This model suggests that the bursting phenomenon maybe similar to the earthquakes, and like them it might be governed by self organized criticality (SOC); indeed SOC behavior in bursting was observed in several magnetars (G{\"o}{\v g}{\"u}{\c s} et al. 1999; 2000, Gavriil et al. 2004, Scholz \& Kaspi 2011), lending support to the crust fracturing scenario. An alternative mechanism for bursts, again in the presence of extremely strong magnetic fields, is magnetic reconnection (Lyutikov 2003, 2015). In a simplified way, the scales of fracturing or reconnection (or even the combination of both processes) are reflected in the energetics of bursts (Thompson and Duncan 2001, Lyutikov 2015). Moreover, bursting activity sometimes affects radiative behavior of the source, e.g., long lasting increase of the persistent X-ray flux (Rea \& Esposito 2011).

Contrary to magnetars, the bulk of the neutron star population is powered via loss of their rotational energy and emit radiation as radio pulsars. RPPs have a wide range of surface magnetic fields; young objects characteristically have B-fields of about $10^{12}$ G. Among them, there are about ten currently known systems with inferred surface magnetic strength in excess of $10^{13}$ G, with few as high as the typical magnetar regime (Ng \& Kaspi 2011). It was one of these high B-field sources (PSR\,J1846$-$0258 with $B=4.9\times10^{13}$ G; Gavriil et al. 2008) which was observed to emit magnetar-like X-ray bursts. Interestingly, PSR\,J1846$-$0258 is an X-ray pulsar without observed radio emission. 

\xsrc is a young isolated neutron star among the group of high B-field systems, with a spin period of $P=0.407$\,s, and an inferred surface dipole field strength of $4.1\times$10$^{13}$ G (Camilo et al. 2000). It is a highly energetic rotation powered object (\.{E} is $2.3\times$10$^{36}$ erg/s) which emits pulsed radiation in a wide range of the electromagnetic spectrum including gamma rays (Parent et al. 2011). Another intriguing property of \xsrc is that it exhibited rotating radio transient (RRAT)-like behavior following the 2007 glitch, therefore, it is the only source with glitch induced radiative changes in radio wavelengths (Weltevrede, Johnston, \& Espinoza 2011, Antonopoulou et al. 2015).

The first magnetar-like triggered bursts from PSR J$1119-6127$ were detected with the {\it Fermi}/Gamma-ray Burst Monitor (GBM) on 2016 July 27 (Younes et al. 2016) and with the {\it Swift}/Burst Alert Telescope (BAT) on July 28 (Kennea et al. 2016). These bursts were coincident with some other extraordinary behavior. In particular, its persistent X-ray flux was increased in excess of 160$-$fold, and it underwent another large glitch (Archibald et al. 2016). Additionally, its pulsed radio emission was stopped following the bursts (Burgay et al. 2016a), and reappeared about two weeks later (Burgay et al. 2016b).

We present here the results of our extensive search for additional bursts from \xsrcnos, and the outcomes of our detailed investigations of all identified bursts. Section 2 describes the results of our untriggered burst search in the \textit{Fermi}/GBM and \textit{Swift}/BAT data. In Section 3, we present the results of our detailed spectral and temporal analyses of all bursts and the persistent emission, and in Section 4 we compare the burst properties of \xsrc with those of typical magnetar bursts, and discuss the implications of our results.  

\section{Observations}

The observations described below were obtained with the \textit{Fermi}/GBM and the \textit{Swift}/BAT. The GBM is an all sky monitor on board {\it Fermi} comprising 14 detectors with 8$-$sr field of view. We used GBM time-tagged event (TTE) and CTIME data, which provide data with temporal resolutions of 2 $\mu$s in 128 energy channels and 0.256 s in 8 energy channels, respectively (see Meegan et al. 2009 for a description of the instrument and data types). The BAT is a coded aperture imager with a half-coded field of view of 1.4$-$sr serving as the burst trigger instrument of \textit{Swift} in the $15-150$\,keV energy range. When BAT is triggered by a burst, it records events with a temporal resolution of 100 $\mu$s in 128 energy channels (Barthelmy et al. 2005).

\textit{Fermi}/GBM triggered on a burst on 2016 July 27 (trigger: bn160727543) located within the error box of \xsrc (Younes et al. 2016). Fig. 1 shows the burst light curve in three energy ranges; most of the emission is below 50 keV. Its T$_{90}$ duration based on its photon spectrum\footnotemark \footnotetext{see Lin et al. (2011a) for the description of photon spectrum based T$_{90}$ duration measurement.} is 0.036$\pm$0.009 s, and the duration obtained with a Bayesian blocks technique is $T_{Bayes}=0.040$\,s.  \textit{Swift}/BAT triggered on the next day, 2016 July 28, on another burst (trigger: 706396) also consistent with \xsrc (Kennea et al. 2016). The burst is soft (see Fig. 2) with a Bayesian block duration estimate of 0.186 s. The event was quite faint in the GBM data (see the lower four panels of Fig. 2). The T$_{90}$ duration of this event using the GBM data is 0.240$\pm$0.075 s. 

\begin{figure}
\centering
\vspace{-2cm}
\includegraphics[scale=0.80]{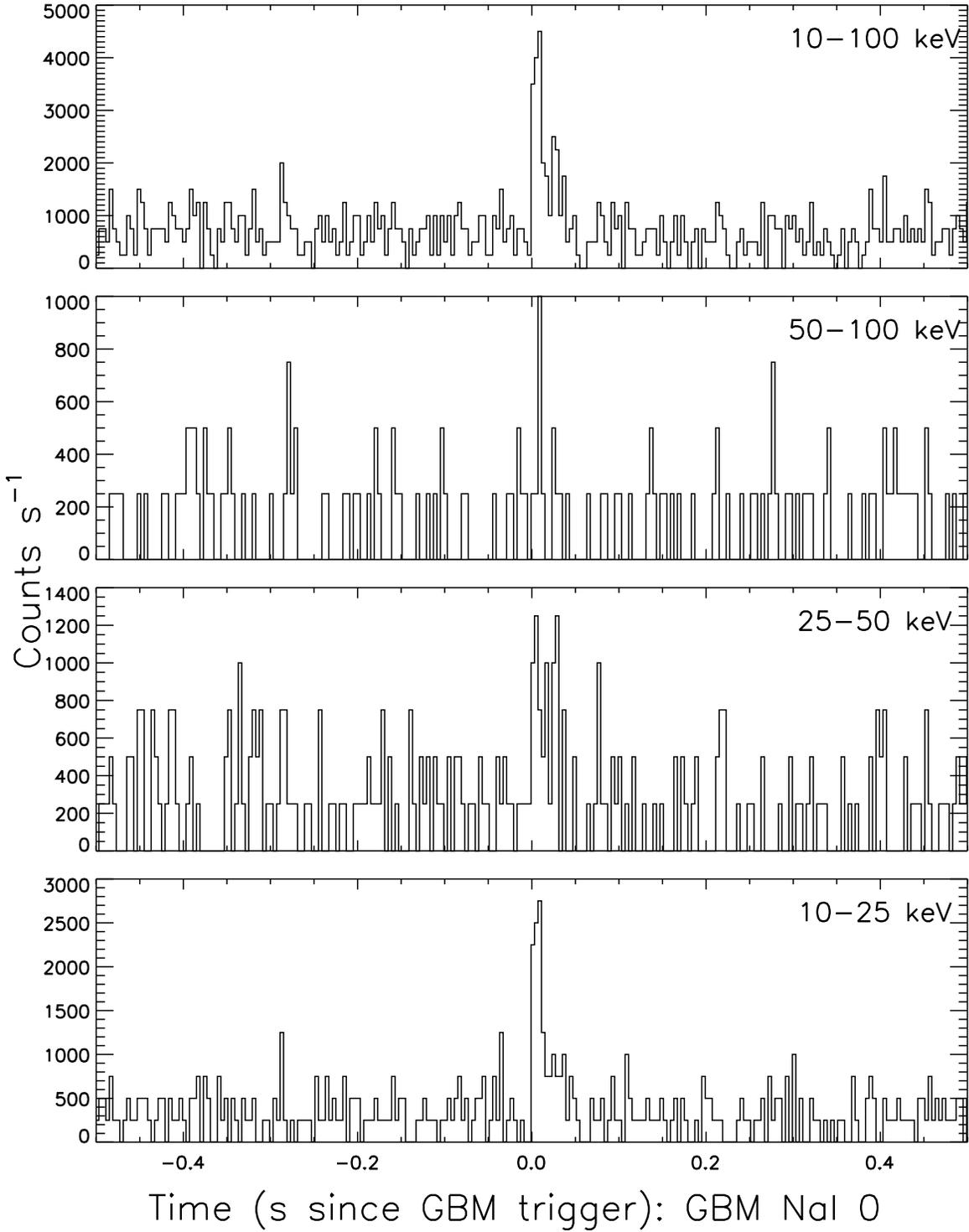}
\vspace{-1cm}
\caption{\textit{Fermi}/GBM light curves of the 2016 July 27 \xsrc burst in three energy ranges as indicated on the panels. The time resolution is 4 ms.} 
\label{gbm27}
\end{figure}

\begin{figure}
\centering
\vspace{-2cm}
\includegraphics[scale=0.80]{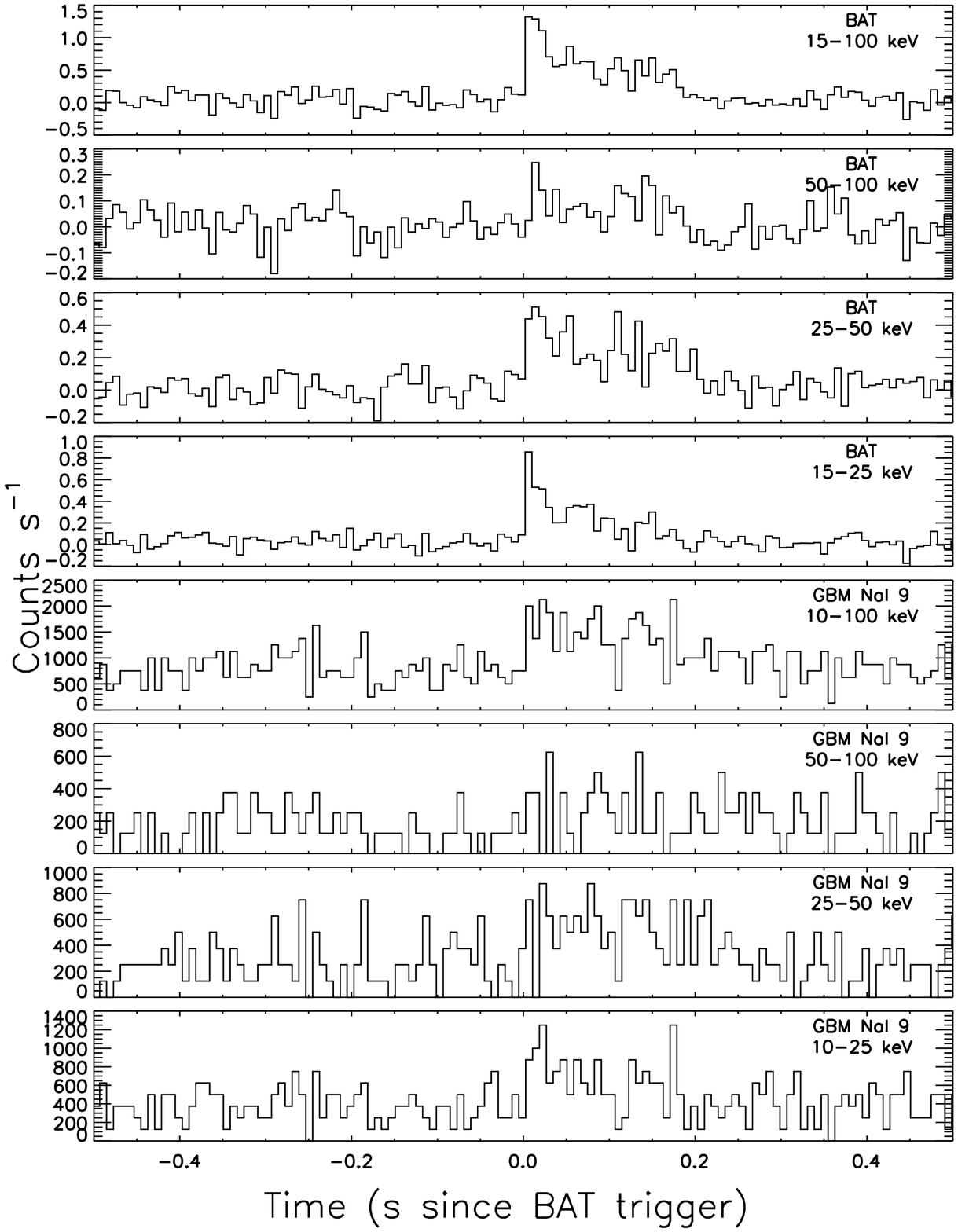}
\vspace{-1cm}
\caption{\textit{Swift}/BAT light curves of the 2016 July 28 burst from \xsrc. The top four panels are obtained with the BAT data in four energy ranges (as indicated on the panels). The lower four panels are obtained with the \textit{Fermi}/GBM data in the energy ranges indicated on the respective panels. All light curves are plotted with 8 ms time resolution.}
\label{gbm28}
\end{figure}

We performed extensive searches in the continuous CTIME and CTTE data of \textit{Fermi}/GBM, as well as in the readout data of the {\it Swift}/BAT trigger to recover bursts which were either weak or could not trigger the instruments for other reasons. We employed two independent search techniques based on a signal-to-noise (S/N) ratio and on Bayesian blocks. Both methods were optimized to search for magnetar bursts (see Kaneko et al. 2010 for the details of the S/N based search, and Lin et al. 2013 for the Bayesian blocks search). Our searches spanned about a week starting on 2016 July 25.

\begin{deluxetable}{ccccccc}
\tabletypesize{\scriptsize}
\tablecaption{\textit{Swift}/BAT and \textit{Fermi}/GBM bursts from \xsrcnos. \label{tab:bursts}}
\tablewidth{0pt}
\tablehead{
\colhead{Burst} & \colhead{Start time*} & \colhead{Instrument} & \colhead{Detection\tablenotemark{**}} & \colhead{$T_{Bayes}$} & \colhead{T$_{90}$} & \colhead{Flux\tablenotemark{\dagger}} \\
\colhead{ID}    & \colhead{(UTC)}       & \colhead{}           & \colhead{Method}    & \colhead{(s)}      & \colhead{(s)}   &\colhead{} \\
}
\startdata
U1	&	2016-07-26 21:15:59.657	&	GBM	&	S/N, BB & 1.456 & 1.8$\pm$0.3 & 0.6$\pm$0.1 \\
U2	&	2016-07-27 12:10:42.325	&	GBM	&	S/N, BB & 0.024 & 0.02$\pm$0.02 & 13.8$\pm$1.3  \\
U3	&	2016-07-27 12:10:53.125	&	GBM	&	S/N, BB & 0.032 & 0.10$\pm$0.05 & 8.3$\pm$0.8	\\
U4	&	2016-07-27 12:19:10.294	&	GBM	& BB & 0.192 & 0.06$\pm$0.07 & 1.1$\pm$0.2	\\
U5	&	2016-07-27 12:17:52.910	&	GBM	&	S/N, BB & 1.000 & 0.8$\pm$0.2 & 1.1$\pm$1.1	\\
T1	&	2016-07-27 13:02:07.872	&	GBM	&	S/N, BB & 0.040 & 0.036$\pm$0.009 & 7.8$\pm$0.7	\\
U6	&	2016-07-27 15:20:21.823	&	GBM	&	S/N, BB & 0.768 & 0.50$\pm$0.3& 1.7$\pm$0.2	\\
U7	&	2016-07-27 15:45:23.156	&	GBM	&	BB & 0.088 & 0.080$\pm$0.03 & 3.8$\pm$0.4	\\
T2	&	2016-07-28 01:27:51.254	&	BAT	& BB & 0.180 &              & 2.4$\pm$0.3	\\
UT2\tablenotemark{\dagger\dagger}	&	2016-07-28 01:27:51.248	&	GBM & S/N, BB	&	0.176 & 0.24$\pm$0.08 & 3.3$\pm$0.5		\\
U8	&	2016-07-28 01:29:27.234	&	BAT	&	BB &  0.020 &             & 3.0$\pm$1.3	\\
U9	&	2016-07-28 01:30:02.462	&	BAT	&	BB & 0.028 &              & 2.5$\pm$0.5	\\
U10	&	2016-07-28 10:47:13.690	&	GBM	&	BB & 0.040 & 0.06$\pm$0.05 & 4.8$\pm$0.6	\\
\enddata
\tablecomments{$^*$The start time of bursts as determined with the Bayesian blocks search \\
$^{**}$BB indicates Bayesian Blocks and S/N indicates Signal over Noise ratio search method \\ 
$^\dagger$ GBM fluxes are in the 8$-$200 keV band, BAT in 15$-$250 keV; both are in units of 10$^{-7}$~erg~cm$^{-2}$~s$^{-1}$ \\ 
$^{\dagger\dagger}$ The burst that triggered BAT.}
\end{deluxetable}

We identified 5 untriggered bursts from \xsrc in the \textit{Fermi}/GBM data using the S/N based search (U1, U2, U3, U5, U6 in Table 1). The burst which triggered \textit{Swift}/BAT was also found in the GBM data but was not bright enough to pass the GBM trigger thresholds (UT2). With the Bayesian blocks algorithm, we identified 5 additional events (U4, U7, U8, U9, U10), for a total of 10 untriggered bursts. Table 1\footnotemark \footnotetext{An expanded version of Table 1, and light curves of all GBM detected events are at http://magnetars.sabanciuniv.edu/psrj1119.php} contains the list and observational details of all \xsrc events observed with BAT and GBM.

\section{Spectral Analysis Results}

\subsection{Bursts}

We fit the time-integrated spectrum of the GBM triggered burst ($8-200$ keV) starting at trigger time and covering a duration of 0.040\,s, using {\it rmfit}\footnotemark \footnotetext{http://fermi.gsfc.nasa.gov/ssc/data/analysis/rmfit/}. The background level was determined by modeling long pre- and post-burst intervals. We used continuum models which best represent magnetar burst spectra: two blackbodies (BB+BB), and the Comptonized model (Compt). We also used simpler continuum models; a blackbody function (BB) and a power law (PL). We find that both BB+BB and Compt represent the spectrum well. The fit with BB+BB yields $kT_1=3.6\pm0.8$ keV and $kT_2=12.3\pm2.1$keV (C Statistics (CStat, Cash 1979)/degrees of freedom (dof) = 176.3/238). Modeling with Compt results in a photon index of $1.0\pm0.6$ and a peak energy of $32.5\pm6.4$ keV (CStat/dof= 177.8/239). The single model fits were worse: we find for the BB temperature, $kT=8.5\pm0.8$ keV (CStat/dof= 191.6/240) and for the PL an index $\gamma=2.2\pm0.1$ (CStat/dof= 186.8/240). The fluence of the burst in the $8-200$ keV band is ($4.1\pm0.4$)$\times10^{-8}$ erg cm$^{-2}$; the corresponding luminosity and total isotropic energy are ($9.3\pm0.8$)$\times10^{39}$ erg s$^{-1}$ and ($3.7\pm0.3$)$\times10^{38}$ erg, respectively, assuming a distance to the source of 8.4 kpc (Caswell et al. 2004).

The burst that triggered \textit{Swift}/BAT was also observed in the CTTE data of \textit{Fermi}/GBM. Therefore, we were able to perform a joint analysis of the two instrument spectra and constrain their parameters better. To this end, we extracted the BAT spectrum in the $15-150$ keV band for the entire 0.18 s burst duration, and a simultaneous GBM spectrum using CTTE data in $8-200$ keV. Applying the same models, we find that the BB+BB model describes the joint spectra best: $kT_1=3.8^{+2.2}_{-1.5}$ keV and $kT_2=11.0^{+1.8}_{-1.0}$ keV ($\chi^2$/dof= 18.1/21). The Compt model fits the joint data but the photon index parameter could not be constrained. The fit with a single BB is also good; $kT=9.7\pm0.6$ keV ($\chi^2$/dof= 23.3/24), while the PL model fit is much poorer ($\chi^2$/dof= 35.4/24). The fluences obtained with the BAT and GBM spectra (15$-$150 keV and 8-200 keV) are ($4.4\pm0.6$)$\times10^{-8}$ and ($6.1\pm0.9$)$\times10^{-8}$ erg cm$^{-2}$, respectively. The burst luminosity and total isotropic energy corresponding to the GBM fluence are ($2.8\pm0.4$)$\times10^{39}$ erg s$^{-1}$ and ($5.2\pm0.8$)$\times10^{38}$ erg, respectively.

The untriggered events from \xsrc have much lower peak intensities, while their emission lasts longer than the triggered bursts. We, therefore, modeled their integrated spectra uniformly with a single BB function, and obtained statistically acceptable results with a BB temperature range between 4.0 and 11.2\,keV. In Table 1, we list their flux values in the 8$-$200 keV and 15$-$150 keV bands for GBM and BAT detections, respectively. Their fluences are between $7\times10^{-9}$ and $1.1\times10^{-7}$ erg cm$^{-2}$, and their corresponding isotropic energies range between 6$\times$10$^{37}$ and 9.3$\times$10$^{38}$ erg, respectively.

\begin{figure}
\centering
\vspace{0cm}
\includegraphics[scale=0.90]{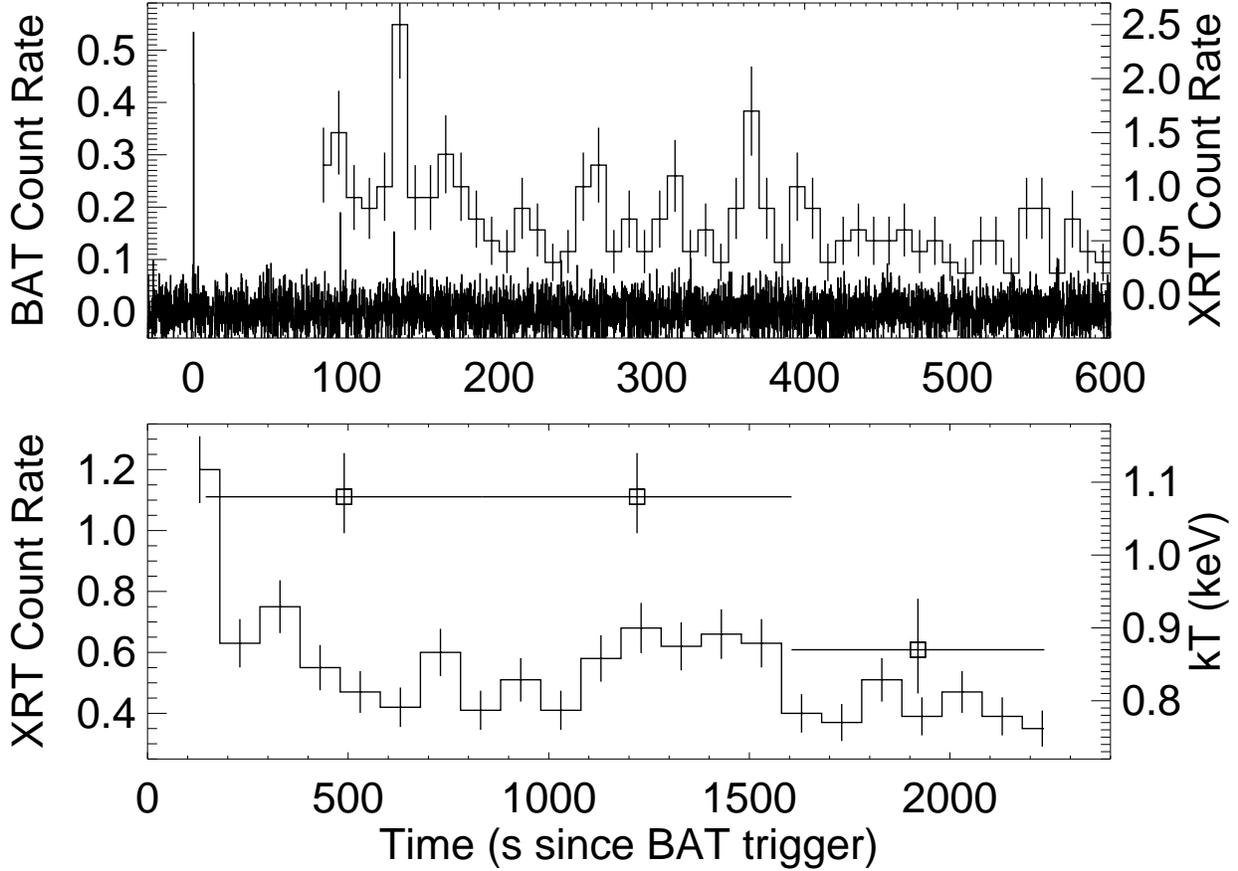}
\caption{(upper panel) \textit{Swift}/BAT light curve of the \xsrc persistent emission ($15-150$\,keV) after the 2016 July 28 burst (left axis) with 4 ms time resolution. The right axis corresponds to the \textit{Swift}/XRT light curve of the source ($0.5-10$\, keV) in 10\,s intervals, starting $\sim100$\,s after the BAT trigger. (lower panel) Extended view of the \textit{Swift}/XRT light curve in the same energy band with 100\,s bins. The squares and the right axis correspond to the BB temperatures of the persistent emission spectral fits as described in the text.}
\label{batxrt}
\end{figure}

\subsection{Enhancement of the Persistent Emission}

The BAT trigger on July 28 was followed with a rapid slew towards PSR J$1119-6127$, and data accumulation with the {\it Swift}/X-Ray Telescope (XRT, Burrows et al. 2004). The XRT observations ($0.5-10$ keV) started $\sim100$\,s after the BAT trigger and lasted for $\sim2.2$\, ks in Photon Counting (PC) mode. We collected events from a circular region of radius $30\arcsec$ centered on the source using the {\it xselect} tool, and after removing the background using a larger circular region of $141\arcsec$ from a source free area, generated the light curve of the source persistent emission in 10\,s time intervals. The upper panel of Fig. 3, displays the XRT light curve of the source, along with the BAT observations. Two untriggered bursts in the data readout of BAT (U8 and U9 in Table 1) have also been seen in XRT. We note an X-ray flux enhancement which declined rapidly, possibly induced by the burst (see the lower panel of Fig. 3). 

To study the spectral evolution of the source during the flux decay, we extracted source spectra ($0.5-10$\,keV) during time intervals corresponding to $140-810$\,s, $810-1580$\,s, and $1580-2250$\,s, after the BAT trigger time. We excluded the first 60\,s of XRT observations to avoid contamination from bursts. We modeled all three spectra simultaneously with a BB function, that is commonly employed for the extended tails of magnetar bursts (see e.g., Lenters et al. 2003). The fit yields a common hydrogen absorption column density of $N_{\rm H}=$($1.13\pm0.15$)$\times10^{22}$ cm$^{-2}$ ($\chi^2$/dof= 64.3/84). This column density is in perfect agreement with the Galactic value in the direction of the source. We find that the BB temperature of the first two segments were consistent with each other; therefore, we linked the two temperatures and repeated the fit. We find that the BB temperature in the first two segments was $1.08\pm0.05$\,keV and decayed to $0.87\pm0.06$\,keV in the third (see lower panel of Fig. 3). The inferred radius of the BB emitting region remains constant (within errors), 1.6$\pm$0.2 km. To determine the longer term temperature evolution of the persistent emission, we accumulated spectrum from the following \textit{Swift}/XRT pointing (Observation ID: 00034632001 with total exposure of 10\,ks, spanning 57\,ks to 92\,ks after the BAT trigger). We find that the spectrum of the persistent emission modeled with a BB (and with N$_{\rm H}$ fixed at $1.13\times10^{22}$ cm$^{-2}$) results in a temperature of $0.87\pm0.01$\,keV, consistent with the temperature obtained during the third segment of the extended tail emission. We also modeled the spectra of the three post-burst segments simultaneously with an absorbed power law model ($N_{\rm H}$ fixed at the same value). The power law model fit is not statistically acceptable ($\chi^2$/dof= 207.2/86); therefore, a non-thermal behavior of the enhanced X-ray emission is ruled out.

\section{Discussion}

\xsrc is an intriguing neutron star in many ways. The latest addition to its extraordinary properties is the emission of short but energetic hard X-ray bursts. We have performed detailed spectral and temporal investigations of the two bursts that triggered \textit{Fermi}/GBM and \textit{Swift}/BAT. We also performed extensive searches for lower luminosity bursts, and uncovered 10 additional events: a total of 12 bursts were detected from \xsrc during its burst active phase of 2016 July 26 to 28. We obtain a cumulative energy of all 12 events as 4.8$\times$10$^{39}$ erg, with an average burst energy of 4$\times$10$^{38}$ erg.  The average burst energy is around the low end of the distribution of short magnetar burst energetics, similar to the average burst energy of 1E 2259+586 (Gavriil et al. 2004).

The two triggered bursts from \xsrcnos, as well as all untriggered events, appear to be typical magnetar bursts\footnote{The bursts from the other low field sources, PSR\,J$1846-0258$ and SGR\,$0418+5729$, were also quite normal.}. Burst durations range from tens of milliseconds to about a second, similar to short bursts from other magnetars (G{\"o}{\v g}{\"u}{\c s} et al. 2001, Gavriil et al. 2004, van der Horst et al. 2012). The spectra of the two triggered bursts are well represented with the Comptonized model, or the sum of two blackbodies with temperatures of about 3 keV and 10 keV, in line with other magnetar bursts (van der Horst et al. 2012, Lin et al. 2011b). The duration of the burst active episode of \xsrc, and clustering of bursts throughout this active phase resemble those of magnetars with low burst rates (G{\"o}{\v g}{\"u}{\c s} 2014). This unusual high field radio pulsar has thus demonstrated typical magnetar behavior. 

We also uncovered a probably burst-induced X-ray intensity increase which lasted about 1400 s. The enhancement is thermal in nature, with evidence of a cooling trend during the tail.  Burst tails with a thermal cooling trend have been seen in other magnetars: SGR 1900+14 (Lenters et al. 2004), SGR 1806$-$20 (G{\"o}{\v g}{\"u}{\c s} et al. 2011), 4U 0142+61 (Gavriil et al. 2011, Chakraborty et al. 2016), and SGR J1550$-$5418 ({\c S}a{\c s}maz Mu{\c s} et al. 2015). These transient enhancements were interpreted as cooling of heat imparted onto or near the neutron star surface. In the other sources, pulsed X-ray intensity was also observed to rise during the extended tail. For \xsrcnos, X-ray observations were performed in a mode with about 2.5 s time resolution (i.e., about 6 times the spin period of the system).  Despite this, there is clearly extra heating associated with the bursts, which may come from an internal mechanism that could also give rise to the glitch (Perna \& Pons 2011, Antonopoulou et al. 2015).

\xsrc is also an exceptional radio pulsar. In 2007, after a Vela$-$like giant glitch ($\Delta\Omega$/$\Omega$$\sim$4$\times$10$^{-6}$), some components of the radio pulse profile started to exhibit erratic RRAT-like behavior that continued for about 3 months (Weltevrede, Johnston \& Espinoza 2011). No associated X-ray activity was reported (Swift, the only X-ray telescope observing the unocculted sky at the time, has a 4$\sigma$ fluence sensitivity of 4$\times$10$^{-8}$ erg cm$^{-2}$ in the 15$-$150 keV band). In its 2016 activation, \xsrc underwent another large glitch with $\Delta\Omega$/$\Omega$$\sim 6\times 10^{-6}$ (Archibald et al. 2016).  However the radio behavior was quite different: pulsed radio emission ceased after the bursts (Burgay et al. 2016a), reappearing two weeks later (Burgay et al. 2016b).  This diversity of glitch-associated magnetospheric behavior, manifested in both radio and gamma-ray emission, is unique.  

The spin recovery after the 2007 glitch was also unusual, with an over-recovery of the spin-down rate that continued to evolve on a timescale of years (Antonopoulou et al. 2015).  These authors considered scenarios that could explain the unusual spin-down evolution.  Superfluid mechanisms include the possibility of vortices moving inwards (Akbal et al. 2015, see below), or variations in the strength of coupling between superfluid and normal components due to heating (see e.g. Haskell \& Antonopoulou 2014). Magnetospheric changes, caused by crustquakes and/or the superfluid dissipation from the glitch, were required to explain the change in radio behavior, but could in principle also explain the subsequent spin evolution.  


It would for example fit quite naturally in the context of the model developed by Akbal et al. (2015) to account for peculiar recovery after the 2007 glitch in \xsrc.  The authors suggested an extension of the standard vortex creep model (Alpar et al. 1984), the most plausible mechanism for Vela type glitches.  In this model a crustquake induces both vortex unpinning (causing the glitch) and the erratic, transient, radio pulse behavior. Akbal et al (2015) estimated the size of an individual plate involved in crust breaking, $D$, in the 2007 glitch, to be about 6 m. If we assume that magnetic stresses were a dominant agent in breaking the crust and initiating the magnetar like bursts in the 2016 outburst, and that some $N$ pieces of crust, each of volume $D^3$ were involved in powering the series of 12 bursts observed, with total energy release $E_{burst}$ = 4.8$\times$10$^{39}$ ergs, $N~D^3$ ($B^2$/8$\pi$)=$E_{burst}$, we obtain the estimate $B_{14}$=2.3$\times$10$^2$ ($D$/6m)$^{-3/2}$ $N^{-1/2}$, where $B_{14}$ is the magnetic field strength in units of 10$^{14}$ G. If we also assume that the 1.6 km radius inferred for the thermal emission covers a single surface layer of broken plates, then $N\approx$(1.6 km/6 m)$^2$, and $B_{14}$$\approx$0.86 is obtained. This means that the local surface magnetic field needed to power the bursts is larger than the inferred dipole magnetic field, but not much stronger than its strength at the pole. However, there is an uncertainty in the volume $N D^3$ where the magnetic energy is released. If the thermal emission radius 1.6 km is larger than the area of the surface at which the crust breaking took place, because of the diffusion of the dissipated energy by thermal conduction or magneto-elastic waves, then $N$ would be smaller and the estimated $B_{14}$ could be larger. 

In summary, the observations of magnetar-like bursts from \xsrc provide the following new insights.  Firstly they provide further evidence that global dipole fields above the quantum critical magnetic field strength are not essential for the magnetar burst trigger mechanism to operate.  Since bursting has not been observed from the majority of radio pulsars it seems clear that there is some minimum field required, however, and this might motivate a detailed X-ray survey of high field radio pulsars to establish the precise threshold for bursting activity. Secondly, \xsrc is the first source to demonstrate such a wide range of behavior associated with glitches and crustal heating: with variation in pulsed radio emission and now the occurrence of bursts.  The superfluid, crust behavior, thermal and magnetospheric properties are an interconnected puzzle, and theoretical models must treat these elements together.  

\acknowledgments

EG and YK acknowledge support from the Scientific and Technological Research Council of Turkey (T\"UB\.ITAK, grant no: 115F463). LL is supported by the Fundamental Research Funds for the Central Universities and the National Natural Science Foundation of China (grant no. 11543004). OJR acknowledges support from Science Foundation Ireland under Grant No.~12/IP/1288.


\begin{thebibliography}{}

\bibitem[Akbal et al.(2015)]{Akbal2015} Akbal, O., G{\"u}gercino{\u g}lu, E., {\c S}a{\c s}maz Mu{\c s}, S., \& Alpar, M.~A.\ 2015, \mnras, 449, 933 

\bibitem[Alpar et al.(1984)]{Alpar1984} Alpar, M.~A., Pines, D., Anderson, P.~W., \& Shaham, J.\ 1984, \apj, 276, 325 

\bibitem[Antonopoulou et al. (2015)]{Antono2015} Antonopoulou, D., et al. 2015, \mnras, 447, 3924

\bibitem[Archibald et al.(2016)]{Archibald2016} Archibald, R.~F., Kaspi, V.~M., Tendulkar, S.~P., \& Scholz, P.\ 2016, \apjl, in press, arXiv:1608.01007

\bibitem[Barthelmy et al.(2005)]{Bart2005} Barthelmy, s., et al. 2005, Space Science Reviews, 120, 143

\bibitem[Burgay et al.(2016a)]{Burgay16a} Burgay, M., et al. 2016a, The Astronomer's Telegram, 9286

\bibitem[Burgay et al.(2016b)]{Burgay16b} Burgay, M., et al. 2016b, The Astronomer's Telegram, 9366

\bibitem[Burrows (2004)]{Burrows2004} Burrows, D.~N. et al., 2004, SPIE, 5165, 201

\bibitem[Camilo et al.(2000)]{Camilo2000} Camilo, F., et al. 2000, \apj, 541, 367

\bibitem[Caswell et al.(2004)]{Caswell2004} Caswell, J.~L., McClure-Griffiths, N.~M., \& Cheung, M.~C.~M.\ 2004, \mnras, 352, 1405 

\bibitem[Chakraborty et al. (2016)]{Chakraborty2016} Chakraborty, M., G{\"o}{\u g}{\"u}{\c s}, E., {\c S}a{\c s}maz Mu{\c s}, S., \& Kaneko, Y.\ 2016, \apj, 819, 153 

\bibitem[Dib \& Kaspi (2014)]{Dib2014} Dib, R., \& Kaspi, V.~M.\ 2014, \apj, 784, 37

\bibitem[G{\"o}{\v g}{\"u}{\c s}  et al.(1999)]{Gogus1999} G{\"o}{\v g}{\"u}{\c s} , E., et al. 1999, \apjl, 526, L93 

\bibitem[G{\"o}{\v g}{\"u}{\c s} et al.(2000)]{Gogus2000} G{\"o}{\v g}{\"u}{\c s}, E., et al. 2000, \apjl, 532, L121 

\bibitem[G{\"o}{\v g}{\"u}{\c s} et al.(2001)]{Gogus2001} G{\"o}{\v g}{\"u}{\c s}, E., et al. 2001, \apj, 558, 228 

\bibitem[G{\"o}{\v g}{\"u}{\c s} et al.(2011)]{Gogus2011} G{\"o}{\v g}{\"u}{\c s}, E., et al.\ 2011, \apj, 740, 55 

\bibitem[G{\"o}{\v g}{\"u}{\c s}(2014)]{Gogus2014} G{\"o}{\v g}{\"u}{\c s}, E.\ 2014, Astronomische Nachrichten, 335, 296

\bibitem[Gavriil et al.(2004)]{Gavriil2004} Gavriil, F.~P., Kaspi, V.~M., \& Woods, P.~M.\ 2004, \apj, 607, 959 

\bibitem[Gavriil et al.(2008)]{Gavriil2008} Gavriil, F.~P., et al. 2008, Science, 319, 1802 

\bibitem[Gavriil et al.(2011)]{Gavriil2011} Gavriil, F.~P., Dib, R., \& Kaspi, V.~M.\ 2011, \apj, 736, 138

\bibitem[Haskell et al.(2014)]{Hask2014} Haskell, B. \& Antonopoulou, D. 2014, \mnras, 438, L16

\bibitem[Kaneko et al.(2010)]{Kaneko2010} Kaneko, Y., G{\"o}{\v g}{\"u}{\c s}, E., Kouveliotou, C., et al.\ 2010, \apj, 710, 1335 

\bibitem[Kennea et al.(2016)]{Kennea2016} Kennea, J. A., et al. 2016, The Astronomer's Telegram, 9274

\bibitem[Lander et al.(2015)]{Lander2015} Lander, S.~K., Andersson, N., Antonopoulou, D., \& Watts, A.~L.\ 2015, \mnras, 449, 2047 

\bibitem[Lenters et al.(2003)]{Lenters2003} Lenters, G.~T., Woods, P.~M., Goupell, J.~E., et al.\ 2003, \apj, 587, 761 

\bibitem[Lin et al.(2011a)]{Lin2011a} Lin, L., Kouveliotou, C., G{\"o}{\v g}{\"u}{\c s}, E., et al.\ 2011a, \apj, 739, 87 

\bibitem[Lin et al.(2011b)]{Lin2011b} Lin, L., Kouveliotou, C., G{\"o}{\v g}{\"u}{\c s}, E., et al.\ 2011b, \apjl, 740, L16

\bibitem[Lin et al.(2013)]{Lin2013} Lin, L., G{\"o}{\v g}{\"u}{\c s}, E., Kaneko, Y., \& Kouveliotou, C.\ 2013, \apj, 778, 105 

\bibitem[Lyutikov(2003)]{Lyutikov2003} Lyutikov, M.\ 2003, \mnras, 346, 540 

\bibitem[Lyutikov(2015)]{Lyutikov2015} Lyutikov, M.\ 2015, \mnras, 447, 1407 

\bibitem[Meegan(2009)]{Maagan09} Meegan, C.\ 2009, \apj, 702, 791

\bibitem[Ng et al.(2011)]{Ng2011} Ng, C.-Y. \& Kaspi, V.~M. \ 2011, in Astrophysics of Neutron Stars, Eds. E. G{\"o}{\v g}{\"u}{\c s}, T. Belloni, \& U. Ertan, AIP Conf. Proc., 1379, 60 

\bibitem[Parent et al.(2011)]{Parent2011} Parent, D., Kerr, M., den Hartog,  P.~R., et al.\ 2011, \apj, 743, 170 

\bibitem[PernaPons (2011)]{PP2011} Perna, R. \& Pons, J.~A. 2011, \apj, 727, L51

\bibitem[Rea et al.(2010)]{Rea2010} Rea, N., Esposito, P., Turolla, R., et al.\ 2010, Science, 330, 944 

\bibitem[Rea(2011)]{Rea2011} Rea, N. \& Esposito, P.\ 2011, Astrophysics and Space Science Proceedings, Volume 21, p.247

\bibitem[{\c S}a{\c s}maz Mu{\c s} et al.(2015)]{SasmazMus2015} {\c S}a{\c s}maz Mu{\c s}, S., G{\"o}{\u g}{\"u}{\c s}, E., Kaneko, Y., Chakraborty, M., \& Ayd{\i}n, B.\ 2015, \apj, 807, 42 

\bibitem[{\c S}a{\c s}maz Mu{\c s} et al.(2014)]{SasmazMus2014} {\c S}a{\c s}maz Mu{\c s}, S., Ayd{\i}n, B., G{\"o}{\u g}{\"u}{\c s}, E.\ 2014, \mnras, 440, 2916

\bibitem[Scholz \& Kapsi(2011)]{SK2011} Scholz, P., \& Kaspi, V.~M.\ 2011, \apj, 739, 94 

\bibitem[Thompson \& Duncan(1995)]{Thompson1995} Thompson, C., \& Duncan, R.~C.\ 1995, \mnras, 275, 255 

\bibitem[Thompson \& Duncan(2001)]{Thompson2001} Thompson, C., \& Duncan, R.~C.\ 2001, \apj, 561, 980 

\bibitem[Turolla et al.(2015)]{TZW2016} Turolla, R., Zane, S., \& Watts, A.~L.\ 2015, Reports on Progress in Physics, 78, 116901

\bibitem[van der Horst et al.(2010)]{vdHorst2010} van der Horst, A.~J., et al. 2010, \apjl, 711, L1 

\bibitem[van der Horst et al.(2012)]{vdHorst2012} van der Horst, A.~J., et al.\ 2012, \apj, 749, 122 

\bibitem[Weltevrede et al.(2011)]{Weltevrede2011} Weltevrede, P., Johnston, S., \& Espinoza, C.~M.\ 2011, \mnras, 411, 1917 

\bibitem[Woods et al.(2004)]{Woods2004} Woods, P.~M., Kaspi, V.~M., Thompson, C., et al.\ 2004, \apj, 605, 378

\bibitem[Younes et al.(2014)]{Younes2014} Younes, G., et al. 2014, \apj, 785, 52

\bibitem[Younes et al.(2016)]{Younes2016} Younes, G., Kouveliotou, C., \& Roberts, O.\ 2016, GRB Coordinates Network, 19736

\end{thebibliography}
\end{document}